\documentclass[iop]{emulateapj}

\usepackage{natbib}

\newcommand{\kms}{km\,s$^{-1}$}
\newcommand{\cii}{[\ion{C}{2}]}
\newcommand{\lfir}{$L_{\mathrm{FIR}}$}
\newcommand{\lcii}{$L_\mathrm{[CII]}$}
\newcommand{\lsun}{$L_\sun$}
\newcommand{\msun}{$M_\sun$}
\newcommand{\msunyr}{$M_\sun$\,yr$^{-1}$}

\newcommand{\mdust}{$M_{\mathrm{dust}}$}

\slugcomment{The Astrophysical Journal Letters, 751:L25 (5pp), 2012 June 1}

\shorttitle{[\ion{C}{2}] and dust emission in a $z=7.1$ quasar host}
\shortauthors{Venemans et al.}

\begin{document}

\title{Detection of atomic carbon [\ion{C}{2}] 158\,$\mu$m and dust emission
  from a $z=7.1$ quasar host galaxy}

\author{B. P. Venemans}
\affil{Max-Planck Institute for Astronomy, K{\"o}nigstuhl 17, 69117
  Heidelberg, Germany}
\affil{European Southern Observatory, Karl-Schwarzschild Strasse 2, 85748
  Garching bei M{\"u}nchen, Germany}
\email{venemans@mpia.de}

\author{R. G. McMahon}
\affil{Institute of Astronomy, University of Cambridge, Madingley Road,
  Cambridge CB3 0HA, UK}
\affil{Kavli Institute for Cosmology, University of Cambridge, Madingley Road,
  Cambridge CB3 0HA, UK}

\author{F. Walter and R. Decarli}
\affil{Max-Planck Institute for Astronomy, K{\"o}nigstuhl 17, 69117
  Heidelberg, Germany}

\author{P. Cox and R.Neri}
\affil{Institut de Radioastronomie Millimetrique, 300 rue de la Piscine, 38406
  Saint Martin d'Heres, France}

\author{P. Hewett}
\affil{Institute of Astronomy, University of Cambridge, Madingley Road,
  Cambridge CB3 0HA, UK}

\author{D. J. Mortlock}
\affil{Astrophysics Group, Imperial College London, Blackett Laboratory,
  Prince Consort Road, London, SW7 2AZ, UK}
\affil{Department of Mathematics, Imperial College London, London, SW7 2AZ,
  UK}

\author{C. Simpson}
\affil{Astrophysics Research Institute, Liverpool John Moores University,
  Twelve Quays House, Egerton Wharf, Birkenhead, CH41 1LD, UK}

\and

\author{S. J. Warren}
\affil{Astrophysics Group, Imperial College London, Blackett Laboratory,
  Prince Consort Road, London, SW7 2AZ, UK}

\begin{abstract}
  Using the IRAM Plateau de Bure Interferometer, we report the detection of
  the 158\,$\mu$m \cii\ emission line and underlying dust continuum in the
  host galaxy of the quasar ULAS J112001.48+064124.3 (hereafter J1120+0641) at
  $z=7.0842\pm0.0004$.  This is the highest redshift detection of the \cii\
  line to date, and allows us to put the first constraints on the physical
  properties of the host galaxy of J1120+0641. The \cii\ line luminosity is
  $1.2\pm0.2\times10^9$\,\lsun, which is a factor $\sim$4 lower than observed
  in a luminous quasar at $z=6.42$ (SDSS J1148+5251). The underlying
  far-infrared (FIR) continuum has a flux density of $0.61\pm0.16$\,mJy,
  similar to the average flux density of $z\sim6$ quasars that were not
  individually detected in the rest-frame FIR. Assuming that the FIR
  luminosity of \lfir\,=$5.8\times10^{11}-1.8\times10^{12}$\,\lsun\ is mainly
  powered by star formation, we derive a star formation rate in the range
  160--440\,\msunyr\ and a total dust mass in the host galaxy of
  $6.7\times10^7-5.7\times10^{8}$\,\msun\ (both numbers have significant
  uncertainties given the unknown nature of dust at these redshifts). The
  \cii\ line width of $\sigma_V=100\pm15$\,\kms\ is among the smallest
  observed when compared to the molecular line widths detected in $z\sim6$
  quasars. Both the \cii\ and dust continuum emission are spatially unresolved
  at the current angular resolution of 2.0$\times$1.7\,arcsec$^2$
  (corresponding to 10$\times$9\,kpc$^2$ at the redshift of J1120+0641). 
\end{abstract}


\section{Introduction}

Understanding the formation of the earliest quasars and host galaxies has
become of great importance since the discovery of the bulge mass--black hole
mass correlation in nearby galaxies, a result which led to the suggestion of a
fundamental relationship between (massive) black holes and spheroidal galaxy
formation \citep[e.g.][]{tre02,mar03}. So far, the stellar component of the
host galaxies of luminous $z\simeq6$ quasars have defied detection at
(observed) optical and near-infrared wavelengths, which is typically attributed
to the central point sources outshining their surroundings. On the other hand,
observations of the molecular gas, through emission lines of, e.g., CO,
redshifted in the millimeter and radio wavebands allows one to probe the total
gas mass and derive the dynamical mass of the high redshift quasar
hosts. These observations indicate that the high-$z$ quasars are associated
with active star forming host galaxies with star formation rates (SFRs)
$\sim$1000\,\msunyr\ \citep{wan08b}.

The C$^+$ fine structure line at 157.74\,$\mu$m (hereafter \cii\ 158\,$\mu$m)
is one of the strongest cooling lines in the warm interstellar medium (ISM),
containing up to 1\% of the far-infrared (FIR) emission in local starburst
galaxies \citep{sta91}. COBE observations of the Milky Way \citep{wri91} show
that this transition is the strongest FIR line emitted by the ISM in our
Galaxy (\lcii\,$=5\times10^7$\,\lsun). For sources at redshifts $z\gtrsim4$,
the \cii\ emission line is redshifted to the sub/millimeter and becomes
observable using ground based facilities. Early attempts to detect the \cii\
line in the $z=4.7$ quasar BR1202--0725 by \citet{isa94} were unsuccessful due
to technological limitations. More recently, \cii\ has been detected in a
number of $z>4$ quasar host galaxies and submillimeter galaxies, including BRI
0952--0115 \citep[$z=4.43$;][]{mai09}, BRI 1335--0417
\citep[$z=4.41$;][]{wag10}, LESS J033229.4 \citep[$z=4.76$;][]{deb11} and SDSS
J1148+5251 \citep[$z=6.42$;][]{mai05}.

To further constrain the build-up of massive galaxies at the earliest cosmic
epochs, it is imperative to locate and study bright quasars at the highest
redshifts possible.  Recently, \citet{mor11} discovered a bright quasar at a
redshift $z = 7.085$, ULAS J112001.48+064124.3 (hereafter J1120+0641), the
only quasar currently known at $z>7$.  The absolute magnitude of $M_{1450}=
-26.6$, the black hole mass of $2.0^{+1.5}_{-0.7}\times10^9$\,\msun\ and
(metal) emission line strengths \citep{mor11} are all comparable to those
observed in $z\simeq6$ quasars. Here we present the detection of the
redshifted \cii\ 158\,$\mu$m emission line and the underlying dust
far-infrared continuum emission in J1120+0641. In Section \ref{observations}
we introduce the observations obtained with the IRAM Plateau de Bure
Interferometer (PdBI), followed by a
description of our results in Section \ref{results}. In Section
\ref{discussion} we discuss the implications of our findings and we present a
summary in Section \ref{summary}. Throughout this Letter a concordance
cosmology with $H_0=70$\,\kms\,Mpc$^{-1}$, $\Omega_{M}=0.3$ and
$\Omega_{\Lambda}=0.7$ is adopted, leading to a luminosity distance for a
source at $z=7.085$ of $D_L=70.0$\,Gpc, a spatial scale of
5.2\,kpc\,arcsec$^{-1}$. The age of the universe at $z=7.085$ is 740\,Myr.

\section{Observations}
\label{observations}

Observations of the \cii\ 157.74\,$\mu$m ($\nu_\mathrm{rest}=1900.54$\,GHz)
line in J1120+0641 at a redshift $z=7.085$ (observed frequency of 235.07\,GHz,
or 1.276\,mm) were carried out between 2011 March and 2012 January with the
WideX correlator using the PdBI. The quasar
was observed on 2011 March 9 for 2.9\,hr with five antennas in C configuration,
and on 2011 December 13 and 15 and 2012 January 12 and 18 for a total of
6.3\,hr with six antennas in D configuration. For the phase calibration the
source J1055+018 was observed every 30 minutes. Absolute flux calibration was
obtained by observing 3C273 and MWC349 before and after each track. The WideX
correlator provides an instantaneous bandwidth of 3.6\,GHz in dual
polarization, which corresponds to $\sim$4400 \kms\ at 235\,GHz. With the goal
to have a better measurement of the continuum level, the frequency setup was
shifted by +1000\,\kms\ after the 2011 March observations. The total
on-source integration time was 8.25\,hr (six antenna equivalent).

The data were reduced using the Grenoble Image and Line Data Analysis
System (GILDAS)
software\footnote{http://www.iram.fr/IRAMFR/GILDAS}. The data were
rebinned to a resolution of 20\,MHz\ (25.5\,\kms). The final
resolution of the image where both frequency setups overlap is
2\farcs02$\times$1\farcs71 at a position angle of 23$^\circ$. 
The rms of the final image is $\sim$1.09\,mJy\,beam$^{-1}$ per 20\,MHz bin.

\section{Results}
\label{results}

\begin{figure*}
\includegraphics[width=\textwidth]{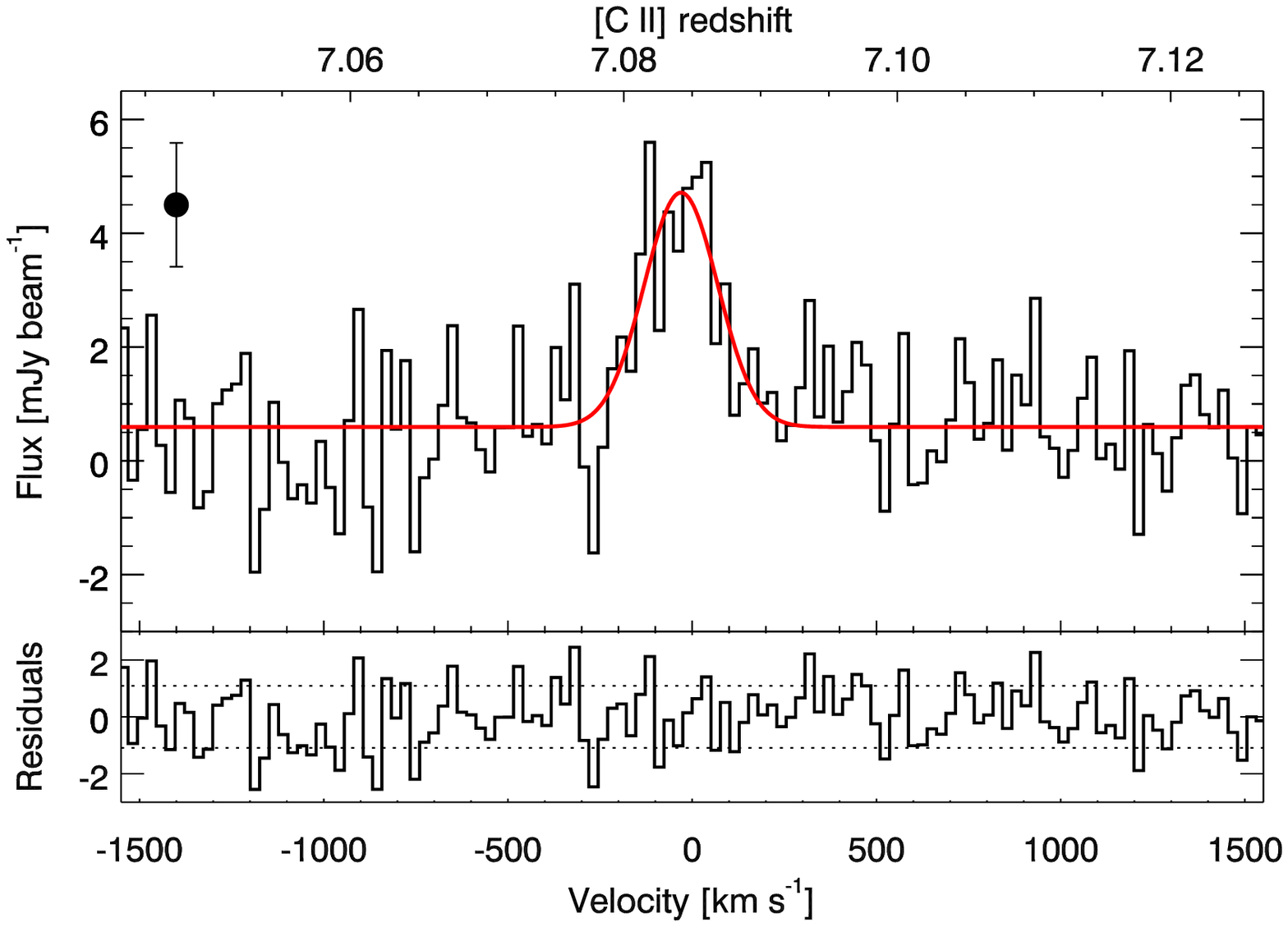}
\caption{Plateau de Bure Interferometer spectrum of J1120+0641 of
  the redshifted \cii\ 158\,$\mu$m line. The channels have a width of
  20\,MHz ($\sim$25\,\kms). The noise per bin is
  1.09\,mJy\,beam$^{-1}$ and is shown in the upper-left corner. The
  published redshift of $z=7.085$ \citep[derived from rest-frame UV
  lines of the quasar;][]{mor11} has been taken as the zero point of the
  velocity scale. The red, solid curve is a Gaussian fit to the
  spectrum and shows that faint continuum emission is also detected in
  the quasar host. The residuals of the fit are plotted below the
  spectrum. The dotted lines represent +$\sigma$ and $-$$\sigma$, with
  $\sigma$ the noise per bin of
  1.09\,mJy\,beam$^{-1}$. \label{spectrum}}
\end{figure*}

The final spectrum of the \cii\ emission line and the underlying continuum
towards J1120+0641 is shown in Figure \ref{spectrum} (top panel). The \cii\
emission is clearly detected at a redshift of $z=7.0842\pm0.0004$. This is
slightly blueshifted (by $-30\pm 14$\,\kms) but consistent with the redshift
of $z=7.085$ \citep{mor11}, which was derived by fitting the template of
\citet{hew10} to the rest-frame UV emission lines. The \cii\ emission line has
a peak flux density of $f_P=4.12\pm0.51$\,mJy\,beam$^{-1}$. In addition, the
underlying (rest-frame) far-infrared continuum is also detected, albeit at
lower significance. Fitting the line with a Gaussian gives a dispersion of
$\sigma_V=99.9\pm14.7$\,\kms, corresponding to a full width at half maximum
(FWHM) of 235$\pm$35\,\kms.

\begin{figure*}
\includegraphics[width=\textwidth]{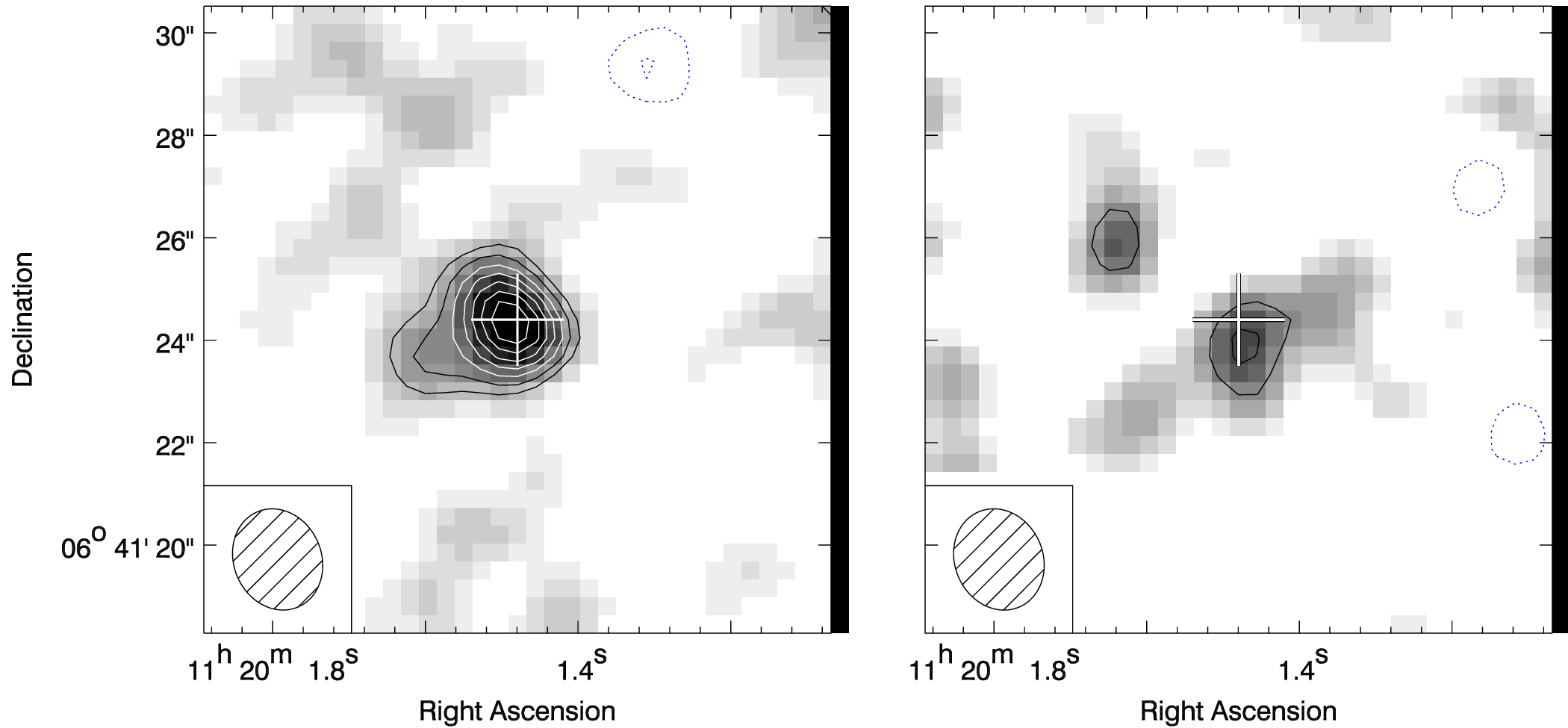}
\caption{Left: image showing the a map constructed from the averaged
  emission between $-$153 and +102\,\kms. The contours are $-$3.5$\sigma$,
  $-$2.5$\sigma$ (blue, dotted lines), 2.5$\sigma$, 3.5$\sigma$ (black, solid
  lines), 4.5$\sigma$, 5.5$\sigma$, 6.5$\sigma$, 7.5$\sigma$ and 8.5$\sigma$
  (white, solid lines), with $\sigma$ the rms noise of
  0.43\,mJy\,beam$^{-1}$. The cross indicates the near-infrared location of
  the quasar. The beam ($2.02$$\times$$1.71$\,arcsec$^2$) is overplotted at
  the bottom left corner of the image -- the emission is unresolved at this
  resolution.  Right: map of the underlying rest-frame FIR continuum
  based on the line-free channels with the same spacing in sigma. The rms
  noise in this map is 0.16\,mJy\,beam$^{-1}$. \label{chmap}}
\end{figure*}

Figure \ref{chmap} shows an image of the \cii\ line and the underlying
continuum emission. The map in the \cii\ emission line (which was averaged
from $-$153\,\kms\ to +102\,\kms) results in a 9.4$\sigma$ detection at the
near-infrared location of the quasar. The integrated flux of the line is
$1.03\pm0.14$\,Jy\,\kms\ (or $8.1\pm1.1\times10^{-21}$\,W\,m$^{-2}$), which
corresponds to a luminosity of \lcii$=1.2\pm0.2\times10^9$\,\lsun. This
\cii\ luminosity is a factor of $\sim$4 lower than that of the
bright $z=6.42$ quasar SDSS J1148+5251
\citep[$4.1\pm0.3\times10^9$\,\lsun;][]{mai05,wal09b}. The width of the \cii\
line of 235$\pm$35\,\kms\ in J1120+0641 is, on the other hand, similar to the
line width seen in SDSS J1148+5251 of 350$\pm$50\,\kms.

The line free part of the spectrum can be used to estimate the far-infrared
continuum emission around 158\,$\mu$m in the rest-frame of the
quasar. Averaging the channels from $-$1500\,\kms\ to $-$290\,\kms\ and from
+215\,\kms\ to +1500\,\kms\ results in a detection of the continuum at a level
of 0.61$\pm$0.16\,mJy. This is a factor $\sim$2 below the average 250\,GHz
flux density found towards all observed $z\sim6$ quasars\footnote{ Although
  the luminosity distance increases by $\sim$20\% between $z=6$ and $z=7.1$,
  the negative $K$-correction results in a flux density at a fixed observed
  frequency that is nearly constant. In our case, the observed continuum at
  235\,GHz will appear $\sim$15\% brighter for a source at $z=7.1$ compared to
  one at $z=6$ observed at 250\,GHz (assuming that the FIR emission comes from a
  grey body with $\beta=1.6$, see Section \ref{firl}).}
\citep[1.26$\pm$0.10\,mJy;][]{wan08b}, but comparable to the average flux
density of $z\sim6$ quasars that were not individually detected at 1\,mm
\citep[0.52$\pm$0.13\,mJy;][]{wan08b}.

Finally, neither the \cii\ emission line nor the underlying continuum are
spatially resolved with the current resolution of the data, i.e.,
$2.02\times1.71$\,arcsec$^2$, corresponding to a spatial scale of
10.4$\times$8.9\,kpc$^2$ at $z\sim7.1$.

\section{Discussion}
\label{discussion}

\subsection{Far-infrared Luminosity}
\label{firl}

The far-infrared luminosity (\lfir) of J1120+0641 can be computed by
approximating the shape of the far-infrared continuum with an optically thin
graybody. Here we assume a dust temperature of $T_d = 47$\,K and dust
emissivity power-law spectral index of $\beta = 1.6$
($\kappa_\nu\propto\nu^\beta$), based on the mean spectral energy distribution
(SED) of quasar hosts in the redshift range $1.8<z<6.4$ \citep[][]{bee06}. The
grey body is scaled to 0.61\,mJy (our continuum detection) at an rest-frame
frequency of $\nu=1900$\,GHz. Following the prescription of \citet{hel88} and
integrating the graybody from 42.5 to 122.5 $\mu$m in the rest-frame, we
derive for J1120+0641 log(\lfir)$=12.15\pm$0.12\,\lsun. Assuming a temperature
of 41\,K and a $\beta=1.95$ \citep[the best-fitting values found for a sample
of $z=4$ quasars,][]{pri01} gives a \lfir\ that is 8\% lower. If we instead
fit our continuum detection to a template of Arp\,220 or M82 \citep{sil98}, we
derive a log(\lfir)$=11.90\pm0.12$\,\lsun\ and
log(\lfir)$=11.97\pm0.12$\,\lsun, respectively. To conclude, our best estimate
for \lfir\ is in the range $5.8\times10^{11}$ -- $1.8\times10^{12}$\,\lsun.

From our detections of the FIR continuum we can estimate an SFR.
A key question here is which fraction of the FIR emission in quasars is
powered by star formation. Studies of lower redshift quasars found fractions
ranging from 25\% \citep[e.g.][]{shi07}, $>$30\%--100\% \citep[e.g.][]{sch06}
to $>$50\% \citep[e.g.][]{wal09b}. Below we assume that all the 158\,$\mu$m
flux density arises from star formation. This should be treated as an upper
limit on the SFR in J1120+0641.

Integrating the grey body from 8\,$\mu$m to 1000\,$\mu$m we get a total
infrared luminosity of $L_\mathrm{TIR}=2.0\pm0.5\times10^{12}$\,\lsun.
Applying the relation between the total infrared luminosity and SFR by
\citet{ken98}, we derive an SFR of $346\pm93$\,\msunyr\ adopting a Salpeter
initial mass function (IMF). This SFR is a factor 1.51 lower when using a
Kroupa IMF. On the other hand, using the template of Arp\,220 (M82), we infer
$L_\mathrm{TIR}=1.3\pm0.3\times10^{12}$\,\lsun\
($L_\mathrm{TIR}=2.0\pm0.5\times10^{12}$\,\lsun) and an SFR of
$219\pm59$\,\msunyr\ ($340\pm91$\,\msunyr).

The measured far-infrared flux density combined with the assumed dust
temperature and emissivity can provide an estimate of the total dust mass
\mdust\ in the quasar host \citep[e.g.,][]{hil83,mag11}:

\begin{equation}
  M_\mathrm{dust} = \frac{f_{\nu,\mathrm{cont}}\,D_L^2}{(1+z)\,\kappa_\lambda\,B_\nu(\lambda,T_d)},
\end{equation}

\noindent
where $f_{\nu,\mathrm{cont}}$ is the far-infrared continuum flux density
measured at rest-frame wavelength $\lambda$, $D_L$ is the luminosity distance,
$B_\nu(\lambda,T_d)$ is the Planck function at $\lambda$ for temperature $T_d$,
and $\kappa_\lambda$ is the dust mass opacity coefficient and is given by
$\kappa_\lambda = 0.77\,(850\,\micron/\lambda)^\beta$\,cm$^2$\,g$^{-1}$
\citep{dun00}. Assuming that the properties of dust grains at $z=7$ are
similar to those at low redshift, we derive \mdust\,$=9\pm2\times10^7$\,\msun,
where the error only includes the uncertainty in the far-infrared continuum
flux.  Alternatively, we can make use of the templates of Arp\,220 and M82 for
which \citet{sil98} derived total dust masses. If the properties of the dust
in J1120+0641 are similar to that in Arp\,220 (M82), then we obtain a dust
mass of $1.3\pm0.3\times10^8$\,\msun\ ($4.5\pm1.2\times10^8$\,\msun).

\subsection{\cii\ Line to FIR Continuum Ratio}
\label{lciilfirratio}

\begin{figure}
\includegraphics[width=\columnwidth]{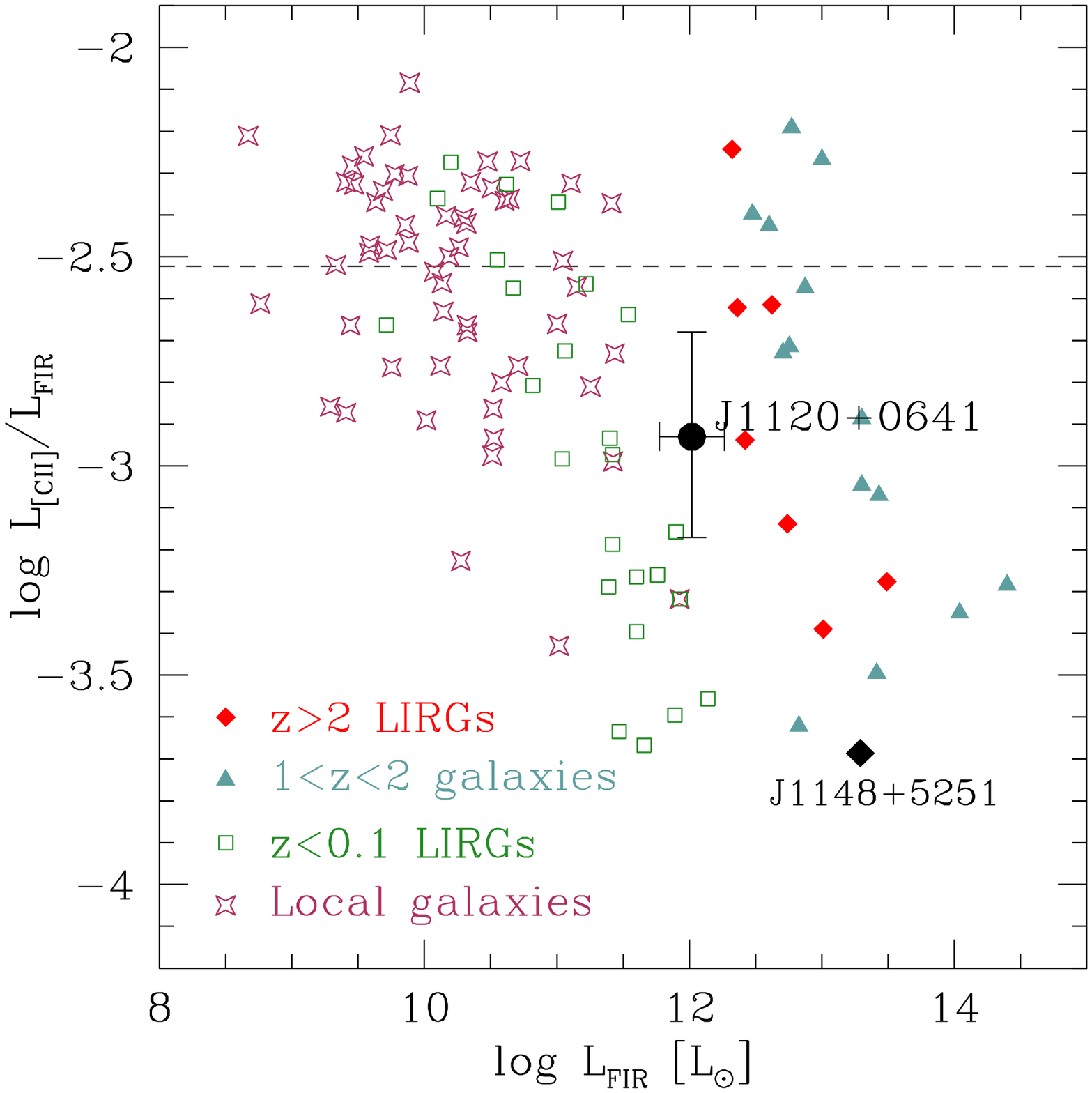}
\caption{Ratio of \cii\ luminosity over FIR luminosity as function of FIR
  luminosity. The values derived for ULAS J1120+0641 are indicated with the
  filled circle. Also plotted are values for local galaxies \citep[open
  stars;][]{mal01}, low redshift luminous infrared galaxies \citep[LIRGs, open
  squares;][]{mai09}, galaxies at $1<z<2$ \citep[filled triangles;][]{sta10}
  and $z>2$ LIRGs \citep[filled
  diamonds;][]{mai09,ivi10,wag10,deb11,cox11}. The \lcii/\lfir\ ratio
  found for J1120+0641 is roughly between the average ratio found for local
  star-forming galaxies (dashed line) and that of the $z=6.4$ quasar
  J1148+5251 (large, black diamond).\label{lciilfir}}
\end{figure}

Surveys of local starbursts and quiescent star-forming galaxies have shown a
tight correlation between \lcii\ and \lfir\ \citep[e.g.,][]{sta91}, although
the ratio \lcii/\lfir\ seems to decrease for galaxies with \lfir\,$>
10^{12}$\,\lsun\ (ULIRGs). In Figure \ref{lciilfir} we plot a compilation of
\lcii/\lfir\ ratios as function of \lfir, obtained from the literature. Also
shown is the measurement of J1120+0641 of
\lcii/\lfir$=6.8\times10^{-4}-2.1\times10^{-3}$, based on our estimate of
\lfir\ of $5.8\times10^{11}$ -- $1.8\times10^{12}$\,\lsun. The ratio we
measure for J1120+0641 is below those of local starburst
galaxies and star-forming galaxies at $z=1-2$ which have \lcii/\lfir\,$\sim
(3.1\pm0.5)\times10^{-3}$. However, it is a factor few higher than the only
other \lcii/\lfir\ ratio currently known above $z>6$, the one measured in the
$z=6.42$ quasar SDSS J1148+5251 \citep[which has
\lcii/\lfir\,$=2.0\pm0.9\times10^{-4}$;][]{mai05}.

\subsection{Dynamical Constraints}

It is instructive to compare the implied gas dynamics to that of other
high-redshift quasars. We here assume that the \cii\ and CO line widths are
roughly the same, which is the case for J1148+5251 at $z=6.42$
\citep{mai05}. The CO lines observed in $z\sim6$ quasars have line FWHMs
between 160 and 860\,\kms, with a median around 350 \kms\
\citep[e.g.][]{wan10,wan11a}. Quasars in the range $1<z<5$ have a very similar
median line width of $\sim$325\,\kms\ \citep{sol05,rie06}. The width of the
\cii\ line in J1120+0641 (235$\pm$35\,\kms) is therefore less than the median
found in $z\sim6$ quasars. Out of 12 published CO line widths in $z\sim6$
quasars only one has a line width less than 235\,\kms\
\citep[FWHM\,=\,160\,\kms\ in SDSS J1044--0125;][]{wan10}.

In the following discussion, we assume that the \cii\ emission comes from a
disk and that we can calculate a dynamical mass from the observed line width
using $M_\mathrm{dyn}\,\mathrm{sin}^2 i
\,\sim\,(\frac{3}{4}\mathrm{FWHM})^2\,R\,/\,G$ \citep{ho07}, where $R$ is the
radius of the source and $i$ its inclination angle. With an observed FWHM of
$235\pm35$\,\kms, we derive a dynamical mass for J1120+0641 of $M_\mathrm{dyn}
=
(7.2\pm2.9)\times10^9\,(R/\mathrm{kpc})\,(\mathrm{sin}\,i)^{-2}$\,\msun. Since
our observations do not resolve the source spatially, we can only obtain an
upper limit on the size of $R\lesssim5$\,kpc. Assuming an (unknown)
inclination angle of 30$^\circ$ leads to a dynamical mass of
$<1.4\times10^{11}$\,\msun. Even in the extreme case that all of this mass was
present in the form of stars in the bulge of the host galaxy and assuming that
the bulge does not extend beyond 5\,kpc, this would imply a black hole-bulge
mass ratio of
$M_\mathrm{BH}/M_\mathrm{bulge}>1.4^{+1.2}_{-0.8}\times10^{-2}$. Unless the
inclination angle is below $\sim$20$^\circ$, this is above the locally
observed ratio of $1.4-3.6\times10^{-3}$ \citep[e.g.,][]{mer01,mar03,gra12},
similar to what has been found in other high-redshift quasars
\citep{wan08b}. Future observations, especially high resolution observations
to derive the size and inclination angle of the emission with ALMA, will
improve the constraints on the dynamical mass of the system and could lead to
a better estimate of the $M_\mathrm{BH}/M_\mathrm{bulge}$ ratio.

\section{Summary}
\label{summary}

Observations at 235\,GHz using the PdBI resulted in the detection of the \cii\
emission line and underlying dust continuum emission in the host galaxy of the
quasar J1120+0641 at $z=7.085$. These results show, for the first time, the
existence of significant amounts of cold gas and dust at $z> 7$, when the
universe was just 740\,Myr old. Based on the mm observations, we can derive the
following properties of the quasar host galaxy:

\begin{itemize}

\item The \cii\ luminosity of $1.2\pm0.2\times10^9$\,\lsun\ is only a factor
  $\sim$4 lower than the one observed in the more luminous $z=6.42$ quasar
  J1148+5251. The width of the line ($235\pm35$\,\kms) is smaller that what
  has typically been found in quasar host galaxies around $z\sim6$. The
  emission remains unresolved at a resolution of 2\farcs02$\times$1\farcs71,
  which corresponds to 10.5$\times$8.9\,kpc$^2$ at the redshift of the line.

\item The far-infrared flux density (at a rest-frame wavelength of
  158\,$\mu$m) of $0.61\pm0.16$\,mJy implies, depending on the model that is
  used, a far-infrared luminosity of
  $5.8\times10^{11}-1.8\times10^{12}$\,\lsun. If all the far-infrared dust
  emission is powered by star formation, the SFR is between 160 and
  440\,\msunyr.  The total dust mass implied by the FIR emission is estimated
  to be in the range $6.7\times10^7-5.7\times10^8$\,\msun.  The \lcii/\lfir\
  ratio of $6.8\times10^{-4}-2.1\times10^{-3}$ is lower than those measured in
  local star-forming galaxies, but (at least) a factor $\sim$3 higher than in
  J1148+5251.

\item If we assume that the \cii\ emission arises from a disk geometry, then
  the narrow line width implies a relatively small dynamical mass of
  $M_{\mathrm{dyn}}=7\times10^9\,(R/\mathrm{kpc})\,(\mathrm{sin}\,i)^{-2}$\,\msun.
  Taking an upper limit on the size of $R<5$\,kpc results in a limit on the
  dynamical mass of $M_{\mathrm{dyn}} <
  3.6\times10^{10}\,(\mathrm{sin}\,i)^{-2}$\,\msun.  Even if all of the
  dynamical mass is in the form of stars, the
  $M_{\mathrm{BH}}$/$M_{\mathrm{bulge}}$ is above the locally observed ratio,
  unless the inclination angle is $i<20^\circ$.

\end{itemize}

Future observations will allow us to better constrain the properties of the
host galaxy of J1120+0641. For example, the detection of the far-infrared
continuum at other frequencies with, e.g., ALMA or the upgraded PdBI, the
Northern Extended Millimeter Array (NOEMA), will constrain the shape of the
FIR SED and thus \lfir. On the other hand, observations of the \cii\ line at
higher spatial resolution will provide information about the extent of the
line emission and will give better constraints on the dynamical mass. Finally,
the detection of other (millimeter) emission lines such as CO,
[\ion{C}{1}]\,370\,$\mu$m, [\ion{C}{1}]\,230\,$\mu$m,
[\ion{O}{1}]\,146\,$\mu$m and/or [\ion{N}{2}]\,122\,$\mu$m (which could be
observed with e.g.\ ALMA or NOEMA) will provide estimates of the total gas
mass and could shed light on the metallicity and ionization state of the ISM
in the host galaxy of J1120+0641.

\acknowledgments We thank the referee for the constructive comments and
suggestions, which improved the manuscript. Based on observations carried out
with the IRAM Plateau de Bure Interferometer. IRAM is supported by INSU/CNRS
(France), MPG (Germany), and IGN (Spain).

{\it Facility:} \facility{IRAM:Interferometer}.

\clearpage

\end{document}